\documentclass{iopart}

\usepackage{amssymb}

\usepackage{amsbsy}

\usepackage{amsmath}

\usepackage{epsfig}

\usepackage{graphicx}

\usepackage{subfig}

\usepackage{iopams}

\eqnobysec

\newcommand{\bea}{\begin{eqnarray}}

\newcommand{\eea}{\end{eqnarray}}

\newcommand{\bn}{\begin{equation}}

\newcommand{\en}{\end{equation}}

\begin{document}

\title {A semi-classical analysis of Dirac fermions in 2+1 dimensions}

\author{Moitri Maiti and R. Shankar}

\address{The Institute of Mathematical Sciences, C.I.T Campus,
Taramani, Chennai-600113, India. }

\eads{\mailto{moitri@imsc.res.in},\mailto{shankar@imsc.res.in}}

%\maketitle

%\date{\today}

\begin{abstract} 
 
We investigate the semiclassical dynamics of massless Dirac fermions in 
2+1 dimensions in the presence of external electromagnetic fields. By 
generalizing  the $\alpha$ matrices by two generators of the $SU(2)$ 
group in the $(2S+1)$ dimensional representation and doing a certain scaling, 
we  formulate a $S\rightarrow\infty$ limit where the orbital and the 
spinor degrees become classical. We solve for the classical trajectories
for a free particle on a cylinder and a particle in a constant magnetic field. 
We compare the semiclassical spectrum, obtained by Bohr-Sommerfeld quantization
with the exact quantum spectrum for low values  of $S$.  For the free particle, the 
semiclassical spectrum is exact. For the particle in a constant magnetic field, 
the semiclassical spectrum reproduces all the qualitative features of the exact 
quantum spectrum at all $S$. The quantitative fit for $S=1/2$ is reasonably 
good.

\end{abstract}

\pacs{73.22.-f, 73.22.Pr, 03.65.Sq}

\noindent

\section{Introduction}

\noindent
The recent developments in discovery of graphene \cite{ref1} 
and topological insulators \cite{ref2} has aroused interest in
the physics of Dirac fermions in two and three spatial dimensions.
In device applications, it is often necessary to consider the system
in the presence of external, spatially varying fields \cite{ref3,ref4}.
While the Dirac equation generally can be solved numerically in these cases,
the semiclassical limit of a quantum system often provides useful 
physical insights into the dynamics. Another interesting aspect of Dirac
fermions is the strong coupling between orbital and spinor dynamics
leading to the well known phenomenon of \textit{Zitterbewegung}. 
This effect in solid-state systems, in particular graphene, has been a topic 
of recent interest \cite{ref5,ref6,ref7}. 

The 2+1 dimensional Dirac equation can be looked upon as the analogue of
the physical 3+1 dimensional Dirac equation. The Hilbert space of Dirac
spinors are then projective unitary irreducible representations of the
2+1 Lorentz group \cite{yip}. In the condensed matter context mentioned
above, there is no Lorentz invariance. The Dirac equation is valid only
in a special frame, namely the rest frame of the material. The physical
significance of the two components of the Dirac spinor depends on the details
of the material as we describe below for graphene and topological
insulators.

The Dirac equation which describes the charge carriers in graphene is a consequence
of graphene's honeycomb lattice crystal structure which consists of two inequivalent 
triangular sublattices. The two bands arise due to the hopping between the two sublattices 
and touch each other at two points, the so called $K$ and $K^{'}$ points of the  Brillouin zone. 
The energy dispersion is linear near these two points. The dynamics of the low energy 
quasi-particles and quasi-holes are effectively described by the massless 2+1 dimensional 
Dirac equation with the role of $c$ played by the Fermi velocity $v_F$. There are two 
species of Dirac fermions, one corresponding to the excitations near the $K$ point and the 
other near the $K^{'}$ point. For each of these, the two components of the Dirac spinor 
represent the probability amplitudes at the two sublattices. While the electrons in graphene are 
confined to two dimensions, they have a physical spin degree of freedom. This manifests
in the effective Dirac theory as an internal degree of freedom. There are thus a total
of four species of Dirac fermions in graphene.

The 2+1 dimensional Dirac equation also describes the low energy quasi-particles
at the surface of 3-dimensional topological insulators \cite{ref8}.
Examples of 3-dimensional topological insulators include $\rm{Bi_2Se_3,
Bi_2Te_3,Sb_2Te_3}$. While in graphene the Dirac equation arises from the honeycomb
lattice structure, in topological insulators it comes from the strong spin-orbit
coupling. Thus unlike graphene where the two components of the Dirac spinor represent
the wave functions on the two sublattices, in topological insulators they represent
the wave functions of the two spin projections. 
 
The earliest approach towards semiclassical analysis of a Dirac particle 
in an electromagnetic field was by W.Pauli\cite{ref9}. 
He determined the phase of a WKB spinor for relativistic point particles 
for some special cases and argued that in the semi-classical limit, the 
translational motion is independent of the spin-degree of freedom. 
This conclusion  was criticized by de Broglie\cite{ref10} with the remark that 
electromagnetic moments are expected to influence
the trajectories in the semi-classical limit. The issue was 
latter addressed and clarified by Rubinow and Keller\cite{ref11}. 
They pointed out that the moments of an electron are 
proportional to $\hbar$ so that in leading order as 
$\hbar\rightarrow0$ the influence of the spin of the particle on the 
trajectories vanishes. Thus in a formalism when the spin is also
taken to be large we can expect it to couple to the orbital
dynamics. We analyze such a formalism in this paper.

In their study of multi-component wavefunction 
Yabana and Horiuchi \cite{ref12} pointed out the necessity of 
including the geometric phase in the Bohr-Sommerfeld quantization condition.
However in their approach and many of the previous approaches to the 
semi-classical analysis of systems with many component wave functions,
\cite{ref13,ref14,ref15} the spin and orbital degrees of freedom are treated 
on a different footing, with the orbital degree of freedom being treated 
semi-classically and spin quantum mechanically. The orbital and spin degrees 
of freedom were combined together for the first time by Pletyukhov 
\textit{et. al}\cite{ref17} in a joint (extended) phase space where the 
description of a spin by continuous variables is achieved by using the basis 
of coherent states. The partition function for a system with 
spin-orbit coupling is then written in the path integral representation 
using coherent states. The semi-classical propagator or trace formula is 
evaluated for the system applying stationary phase approximation\cite{ref16}.

In this paper we follow the extended phase space approach and present a 
systematic formalism for the semi-classical analysis for the 
massless Dirac fermion in 2+1 dimensions in presence of external 
electromagnetic fields. We use the fact that, in 2+1 dimensions, the 3 Dirac 
matrices are elements of the two dimensional representation of the 
$SU(2)$ algebra to generalise them to be the corresponding elements in the $2S+1$
dimensional algebra. We then establish, using the path integral formalism
that there is a systematic stationary phase approximation with $1/S$ being the 
small parameter.
The classical equations of motion couples orbital and the spinor degrees of freedom.
For cases of free particle on a cylinder and particle in a
constant magnetic field closed orbit solutions are found and  
quantized using Bohr-Sommerfeld quantization. 
The classical solutions for the free particle are oscillatory rather
than straight lines thus illustrating the phenomenon of \textit{zitterbewegung}.
Comparison with the exact quantum mechanical results shows that all
the qualitative features of the spectrum are reproduced by the semi-classical
analysis. The quantitative agreement is good even for small values of $S$, 
in particular $S= \frac{1}{2}$.

The organization of the rest of this paper is as follows. In section \ref{SC}, 
we give the formulation of the path integral description of the propagator 
using coherent states and formulate a systematic semiclassical expansion in 
$1/S$. We write down the classical equations of motion and the quantization 
condition for the closed classical orbits.
In section \ref{free}, the Dirac fermion on a cylinder in absence of any 
external field is studied. The particle in a constant external magnetic field 
is studied in section \ref{mag}. The semi-classical and exact spectrum are 
compared in section \ref{discussion}. The results are summarized in section \ref{results}.

\section{The Semi-Classical Expansion}
\label{SC}
We begin by outlining the general formalism to analyze the 
spectrum semi-classically. The hamiltonian for a single 
massless $S=\frac{1}{2}$ Dirac particle in 2 spatial dimensions,
in the presence of external magnetic and spatially varying fields 
$V(\mathbf{x})$ is given by:
\bn
\mathcal{H}=v_F\alpha.\Pi +e V(\mathbf{x})\label{hsc1}
\en
where, $\Pi_i=p_i-eA_i$ is the mechanical momentum, $v_F$ is the Fermi velocity. 
The $\alpha$ matrices can be represented by the Pauli matrices and the 
wavefunctions by a two component spinor. 

The hamiltonian in eqn.[\ref{hsc1}] is generalized by replacing the 
$\alpha^x$ and $\alpha^y$ matrices by two of the generators of the 
$SU(2)$ group in the $(2S + 1)$ dimensional representation and written as:
\bn
\mathcal{H}=\frac{v_F}{S} \mathbf{S.\Pi} + e V(\mathbf{x})\label{hsc2}
\en
We note that while the system is rotationally invariant at all $S$, it is not 
Lorentz invariant for $S\ne 1/2$. We will explicitly see the consequences
of this in the spectrum in later sections. However for applications to
condensed matter systems, Lorentz invariance is not crucial. 
We resort to a path integral formalism to describe the time evolution 
of the system and find a systematic semiclassical parameter $S$ for the combined 
orbital and spinor degree of freedom as described below. 

The path integral is constructed using coherent state representation. 
The coherent states which describe the system are direct product 
of the usual phase-space coherent states $\left.|\vec{x},\vec{p}\right\rangle$ and 
$SU(2)$-coherent states 
$\left.|\vec{n},\mu\right\rangle$ i.e
\bn
\left.|\vec{x},\vec{p};\vec{n},\mu\right\rangle=\left.|\vec{x},\vec{p}\right\rangle \otimes \left.|\vec{n},\mu\right\rangle
\en

where  $\vec{n}.\vec{n}+\mu^{2}=1$.
%The spin-coherent states are described by $\left.|\hat{n}\right\rangle=\left.|\vec{n}
%,\mu\right\rangle$ where $\vec{n}$ is a vector in the 2 dimensional plane and $\mu$ is the 
%component of $\hat{n}$ in the orthogonal direction to $\vec{n}$ and $\vec{n}.\vec{n}+\mu^{2}=1$. 

\noindent
The path integral representation of time evolution of the system from time $t_i$ to time $t_f$ 
is:
\bea
\left\langle \vec{x}_f,\vec{p}_f;\vec{n}_f,\mu_f\right.|e^{-\frac{i/ \mathcal{H}(t_i-t_f)}{\hbar}}
|\left.\vec{x}_i,\vec{p}_i;\vec{n}_i,\mu_i\right\rangle
=  \int\mathcal{D}[\vec{n}(\tau,\zeta)]\mathcal{D}[\mu(\tau,\zeta)] \nonumber\\ \mathcal{D}[\vec{x}(t)]\mathcal{D}[\vec{p}(t)]e^{-\frac{i\mathcal{S}}{\hbar}}
\label{tran_amp}
\eea
\bn
\mathcal{S}[\vec{x(t)}; \vec{p(t)}; \vec{n}(\tau,\zeta),\mu(\tau,\zeta)] = 
\int_{t_i}^{t_f}dt(\vec{p}.\frac{d\vec{x}}{dt}-\mathcal{H}) - \hbar S\Omega[\vec{n}(\tau,\zeta),\mu(\tau,\zeta)
]\label{action1}
\en
where $\mathcal{S}$ is the action with:

\bea
\vec{x}(t_i) = \vec{x}_i;\;\; \vec{p}(t_i) = \vec{p}_i;\;\; \vec{n}(\tau_i) = \vec{n}_i; \;\;\mu(\tau_i)=\mu_i \\
\vec{x}(t_f ) = \vec{x}_f;\;\; \vec{p}(t_f ) = \vec{p}_f;\;\; \vec{n}(\tau_f) = \vec{n}_f;\;\;\mu(\tau_f)=\mu_f
\eea
$\Omega[\vec{n}(\tau,\zeta),\mu(\tau,\zeta)]$ is the geometric phase encompassed during time evolution 
and is written as: 
\bn
%\Omega[\hat{n(\tau,\zeta)}]=\int_{\tau_i}^{\tau_f}d\tau\int_0^{1}d\zeta \; \hat{n}.
%\partial_{\tau} \hat{n}\times \partial_{\zeta} \hat{n}\label{sangle}
\Omega[\vec{n}(\tau,\zeta),\mu(\tau,\zeta)]=\int_{\tau_i}^{\tau_f}d\tau\int_0^{1}d\zeta \; 
\mu \epsilon_{ij} \partial_{\tau}n^{i}\partial_{\zeta}n^{j},\;\;\;\;i,j=1,2\label{sangle}
\en
$\vec{n}(\tau;\zeta)$ parameterizes the surface on the unit sphere in 2 dimensions 
defined by the curve $\vec{n}(\tau)$,
and the geodesics from $\vec{n}(\tau_i)$ and $\vec{n}(\tau_f )$ to the 
north pole (or some standard point on the unit sphere). 
For any arbitrary $S$ it is convenient to write $\vec{S}= S\vec{n}$.
The system variables are scaled in terms of the 
characteristic length scale $\lambda$ and time scale $\omega$ 
of the system as:
\bea
& & x_i^{'} = \frac{x_i}{\lambda},\;\; p_i^{'} = \frac{\lambda}{\hbar S} p_i,\;\;t^{'} = \omega t \nonumber\\
& & A_i^{'} = \frac{e \lambda}{\hbar S} A_i,\;\;\Pi_i^{'} = \frac{\lambda}{\hbar S} 
\Pi_i,\;\;V^{'}(x_i^{'})= \frac{1}{e \omega \hbar S}V(x_i)\label{scale1} 
\eea
Using Eqn.[\ref{scale1}] the action $\mathcal{S}$ can be written as:
\bea& &
\mathcal{S}[\vec{x^{'}(t^{'})}; \vec{p^{'}(t^{'})}; \vec{n}(\tau,\zeta),\mu(\tau,\zeta)]\nonumber\\
&=& \hbar S\left[ \left.\int_{t_i}^{t_f}dt^{'}(\vec{p}^{'}.\frac{d\vec{x}^{'}}{dt^{'}}-\mathbb{H})
- \Omega[\vec{n}(\tau,\zeta),\mu(\tau,\zeta)]\right. \right]\label{action1a}
\eea
where, 
\bn
\mathbb{H}=\vec{n}.\Pi^{'}+ V^{'}(x^{'})
\en
Henceforth we will drop the subscript in the variables keeping 
in mind that they are scaled and hence are dimensionless. 
The propagator in eqn.(\ref{tran_amp}) can then be evaluated by a systematic 
stationary phase approximation with 
$1/S$ as the expansion parameter. 
In this paper we will find closed orbit solutions
to the classical equations of motion and quantize them using the 
Bohr-Sommerfeld quantization procedure. 

In the classical limit
the system observables satisfy the 
following Poisson bracket relations:
\bn
\{x_i,p_j\} = \delta_{ij};\;\;\;  \{ n_i,n_j\} = \epsilon_{ij}\mu; \;\;\;  
\{ \mu,n_i\} = \epsilon_{ji}n_j \;\;\; \{n_i,x_j\}= 0 \nonumber\\
\en
and the equations of motion are:
\bea
\dot{x}_i &=&  n_i \nonumber \\ 
\dot{\Pi}_i &=&  \epsilon_{ji}n_j -\partial_i V(x_i),\;\;\;i=1,2\nonumber\\
%\dot{n}_i &=& \epsilon_{ijk}n_k\Pi_j,\;\;\; i=1,2,3
\dot{n}_i &=& \mu \epsilon_{ij}\Pi_j,\;\;\;i,j=1,2\nonumber\\
\dot{\mu} &=& \epsilon_{ij}n_i\Pi_j,\;\;\; i,j=1,2
\eea
If the classical trajectories are closed then the action 
is quantized using Bohr-Sommerfeld quantization condition 
to find out the allowed orbits and subsequently quantize the spectrum. 
In terms of the canonical co-ordinate and the momentum, 
the quantization condition  
encloses an area in the phase space. For the Dirac particle in 
(2+1) dimensions, the quantization condition 
includes the contribution from geometric phase which is otherwise absent for quantization 
of Schr$\rm{\ddot{o}}$dinger particle, as follows:
\bn
\oint_C p_1dx+\oint_Cp_2dy- \Omega[\vec{n},\mu]= 2\pi\frac{m}{S} , \;\; m=0,1,2,\cdots \label{qnuant1}
\en
where $C$ denotes the closed orbits in the phase space.   
We use this method to 
analyze the spectrum for two solvable cases of (i) a free particle 
and (ii) a particle in a constant external magnetic field 
in following sections (\ref{free}) and (\ref{mag}) and compare with the exact solutions.
%\newpage

\section{Free Particle}
\label{free}
\subsection{Semi-classical expansion}
\label{SCfree}
\noindent
Consider a massless Dirac particle on the surface of a cylinder of
circumference $L$. 
In the absence of any spatially varying field 
$V(x)$ and magnetic field, the hamiltonian for the 
system is:
\bn
\mathbb{H}_{free}=\mathbf{n.p}\label{hamil_free}
\en
We take $\lambda =L$ and $\omega=\frac{v_F}{L}$. 
The action is:
\bn
\mathcal{S}[\vec{x}(t); \vec{p}(t); \vec{n},\mu] =  \hbar S \;\left[ 
\int_{t_i}^{t_f}dt[p_x\frac{dx}{dt}-n_i p_i]-\Omega[\vec{n},\mu]\right]
\en
The equations of motion are as follows:
\bea
\dot{x}_i &=&  n_i \;\;, i=1,2\nonumber\\ 
\dot{p}_i &=&  0 \nonumber\\
%\dot{n}_i &=& \epsilon_{ijk}n_k\Pi_j,\;\;\; i=1,2,3
\dot{n}_i &=& \mu \epsilon_{ij}p_j,\;\;\;i,j=1,2\nonumber\\
\dot{\mu} &=& \epsilon_{ij}n_i p_j,\;\;\; i,j=1,2
\eea
We look into solutions for which  $p(t)\;=\;p_x^{0},\;\;p_y(t)=0$ and 
the solutions are:
\bea
x(t) &=& n_x^0 t +  x_0 \nonumber\\
y(t)&=& \frac{1}{\omega}(\mu^{0} \cos \omega t + n_y^{0} \sin \omega t) \nonumber \\
p(t)&=&p_x^0 \nonumber\\
n_x(t) &=& n_x^{0} \nonumber\\
n_y(t) &=& n_y^{0} \cos \omega t - \mu^{0} \sin \omega t \nonumber\\
\mu(t) &=& \mu^{0} \cos \omega t + n_y^{0} \sin \omega t 
\eea
where, $\omega = p_x^{0} \label{omega} $.

The velocity and hence the classical trajectories thus oscillate in the direction transverse to
the momentum illustrating the phenomenon of \textit{zitterbewegung}.
%The hamiltonian acts as a pseudo external field and will 
%try to orient the spin polarization vector along the direction of the 
%momentum. 
%The spin polarization thus precesses around the momentum.
%Thus the velocity and hence the trajectory oscillates in the direction
%transverse to the momentum.
 
The geometrical phase traced out by the closed classical trajectories is:
\bea
\Omega = 2 \pi (1-\cos(\rm{\theta_0}))\label{sangle1}
\eea
where, 
\bn
\cos (\rm{\theta_0}) = \hat{n}.\frac{p_x^{0}\hat{x}}{\omega} = n_x^0 \label{theta}
\en
From commensurate time-period for the linear motion and that for the orbital motion:
\bn
 \frac{2 \pi}{\omega}=\frac{1}{n_x^0} 
\en

The Bohr-Sommerfeld quantization condition following eqn.(\ref{qnuant1}) 
for the free particle is:
\bn
\oint p_x^0 dx - 2\pi (1-\cos \theta)= 2 \pi \frac{m}{S}  
\en
where\; m\; is\; an\; integer.
Substituting the value of $\cos \theta$ from (\ref{theta}), we have:
\bn
p_x^0 - 2 \pi(1-n_x^0) = 2 \pi \frac{m}{S}
\en
Quantizing each of the term individually, we get:
\bea
p_x^0 &=& 2 \pi n/S \\ n_x^0 &=&  (m_1/S + 1) \;\;\;\\ 
\rm{where}\;\; n &=&0,1,2,\cdots,\;\;\;-2S<m_1<0 \nonumber
\eea
The energy of the particle is given by:
\bea
E &=& n_x p_x^0 \nonumber \\
  &=& 2 \pi n(\frac{m}{S})\;\;\; \rm{where}\;\;m=-S,\dots,S\label{esc_free}
\eea

%Restoring the scaling factors,

%\bn

%E=\hbar v_F m\frac{2\pi n}{L}\label{esc_free2}

%\en
Next we analyze the quantum mechanical spectrum of the free Dirac particle 
and compare with the spectrum obtained above. 

\subsection{Quantum spectrum}
\label{QFree}
The hamiltonian in dimensionless units is given by:
\bn
\mathcal{H}=\frac{1}{S} \bf{S.p} \label{h2}
\en
This hamiltonian can be easily diagonalised. 
To compare with the semiclassical spectrum, we consider plane waves
propagating in the $x$ direction. The periodic boundary conditions imply 
that $k_x=2\pi n/L$. $S_x$ has eigenvalues $m=-S,\dots S$.
\bn
%E=\hbar v_F \frac{m}{S}(\frac{2\pi n}{L}) \label{efreeS} 
E=2 \pi n \frac{m}{S}, \;\;\; n=0,1,2 \cdots \label{efreeS} 
\en
% $\rm{where} \; -S < m < S$. 
Comparing equations (\ref{esc_free}),(\ref{efreeS}) we find that the spectrum 
of the free particle obtained semi-classically is exactly identical to its 
quantum mechanical one. 

Also note that the quantum spectrum consists of $2S+1$ massless particles,
travelling with speed $mv_F$. This illustrates our comment earlier that 
except at $S=1/2$, the theory is not Lorentz invariant.

\section{External magnetic field}
\label{mag}
\subsection{Semi-classical quantization}
\label{SCmag}
\noindent
In this section we study the massless Dirac particle in a constant
magnetic field. 
The system is characterized by a length scale,
the cyclotron radius $l_c$ and a time scale, the cyclotron 
frequency $\omega_{c}$ which are:
\bn 
l_c \equiv \sqrt{\frac{\hbar}{eB}} \;\;\; \omega_c \equiv \frac{v_F}{l_c} \label{lcwc}\\
\en
We choose $\lambda=l_c$ and $\omega=\omega_c$. 
The system variables are scaled in terms of the characteristic scales of the system 
following eqn.(\ref{scale1}).
The system is described by: 
\bn
\mathbb{H}_{mag}={\bf{n.\Pi}} 
\en
where $\Pi$ related to the canonical momentum and 
the vector potential as $\Pi_i = p_i + A_i$. 
For a constant magnetic field, $\{\Pi_i,\Pi_j\}=\epsilon_{ij}$. 
We define another observable, the guiding center co-ordinate 
of the system $R_i \equiv x_i+\epsilon_{ij}\Pi_j$,  
which satisfies: 
$
\{R_i,R_j\}=-\epsilon_{ij},\;\;
\{R_i,\Pi_j\}=0\; \& \;\{R_i,n_j\}=0.$
%%\eea
It then follows that $R_i$ are constants of motion, 
\bn
\{ R_i,\mathbb{H}_{mag} \}=0\;\;,i= 1,2 \label{Rcomm}
\en
The canonical momentum $p_i$ and the canonical co-ordinate $x_i$ 
can be written in terms of the guiding
center co-ordinate and the kinetic momentum as follows:
\bea
& &x = R_x-\Pi_y ; \;\;y = R_y+\Pi_x \nonumber\\
& &p_x = (\Pi_x-R_y)/2 ; \;\;p_y = (\Pi_y+R_x)/2 \label{v2}
\eea
The equations of motion of the system in terms of 
$\mathbf{R}\;$, $\mathbf{\Pi}\;$ ,$\mathbf{n}$ and $\mu$ can be written as:
\bea
\dot{R}_i &=&  0\nonumber\\
\dot{\Pi}_i &=& \epsilon_{ji}n_j \nonumber\\
%\dot{n}_i &=& \epsilon_{ijk}n_k\Pi_j,\;\;\; i=1,2,3\label{eom}
\dot{n}_i &=& \mu \epsilon_{ij}\Pi_j,\;\;\;i,j=1,2\nonumber\\
\dot{\mu} &=& \epsilon_{ij}n_i\Pi_j,\;\;\; i,j=1,2\label{eom}
\eea
We choose $R_i(t)=R_i$. Since $R_i$ commutes with the 
hamiltonian it represents the infinite degeneracy of each Landau level 
in the exact solution.
The system has rotation symmetry in the $x-y$ plane and for this closed system 
 the angular momentum is conserved and is given as: 
\bea
\frac{J_z}{S} = \epsilon_{ij}x_ip_j+\mu
              =\frac{1}{2}\Pi_i\Pi_i + \mu - \frac{1}{2}R_iR_i\label{ang_sc}
\eea
in terms of the guiding center co-ordinate and the kinetic momentum. 

The Bohr-Sommerfeld quantization condition %[ eqn. (\ref{qnuant1}) ] 
in terms of $\Pi$ and $R_i$ is given as:
\bn
\frac{1}{2}\oint\left(\epsilon_{ij}\Pi_i\dot{\Pi}_j - \epsilon_{ij}R_i\dot{R}_j \right)dt 
-\Omega[\vec{n},\mu]= 2\pi\frac{m}{S}\label{quant2}
\en
where $\Omega[n,\mu]$ is given by eqn.(\ref{sangle1}). 
To solve for $\mathbf{\Pi}$, $\mathbf{n}$ and $\mu$, we consider the 
following ans\"atze in which a) $\vec{n}  \equiv n_x \hat{x}+ n_y 
\hat{y}$ parallel to $\mathbf{\Pi}$. 
b) $\vec{n}$ is perpendicular to the $\mathbf{\Pi}$. 
We look into each case separately in the following sections. 

\subsubsection{\bf{$\vec{n} \parallel \mathbf{\Pi}$}}
\label{npar}
For this case, we consider $\mu=\cos(\theta)$. 
The components of $\mathbf{\Pi}$ are related to the $azimuthal$ angle $\phi$ in the $x-y$ plane as:
\bea
\Pi_x = \Pi \cos{\phi} ;\;\;\Pi_y = \Pi \sin{\phi}
\eea
The equations of motion (\ref{eom}) in terms of $\phi$ and $\theta$ becomes:
\bea
\sin{\theta}\dot{\rm{\theta}}(t) &=& 0 ;\;\;\cos{\theta}\dot{\rm{\theta}}(t) = 0; \nonumber \;\;
\dot{\Pi}(t) = 0 \nonumber \\
\sin{\theta}\dot{\phi}&=&-\cos{\theta}\Pi;\;\;\Pi\dot{\rm{\phi}}(t) = 2\sin{\theta}  \label{eomscmag1}
\eea
and the solutions are: 
\bea
\theta(t) &=& \theta_0 \\
\Pi(t) &=& \Pi_0\\
\phi(t)&=& \sqrt{\cos(\theta_0)}\;\; t
\eea
where $\cos{\theta_0}\;=\;\mu_0$. The corresponding trajectories are:
\bea
x(t)-R_x &=& -\Pi_0\sin{\phi(t)} \nonumber \\
y(t)-R_y &=&  \Pi_0\cos{\phi(t)} 
\eea
where $\Pi_0$ is related to $\theta_0$ as:
\bn
\label{pi0ans}
\Pi_0^2 = - \frac{\sin{\theta_0}^2}{\cos{\theta_0}},\;\;\; \pi/2<\theta_0<\pi
\en
The particle trajectories are circular orbits centered around the 
guiding center.

For a given $\theta$, the tip of the unit vector thus moves 
in closed orbits. % (eqns(\ref{eomscmag1})). 
The solid angle subtended by the closed path is given by eqn.(\ref{sangle1}). 
We then use the Bohr-Sommerfeld quantization rule  
to pick the allowed orbits. The quantization condition is: 
%The quantization condition in eqn.(\ref{quant2}) using eqn. (\ref{sangle1}) reads:
\bn
\left(\frac{\Pi_0^2}{2}+\cos{\theta_0}\right) =  \frac{m^{'}}{S}+1, \label{quant3}
\en where $m^{'}=-2S, -2S+1, \cdots$. With a change of variable $n=m^{'}+2S$, where $n$ is an integer
 and little algebra we get,
\bn
\label{costhans}
\cos{\theta_0}=\frac{1}{3}
\left( (\frac{n}{S}-1)-\sqrt{(\frac{n}{S}-1)^2+3}\right)
\en
from equations (\ref{pi0ans}) and (\ref{quant3}). 
The energy of the particle is given by:
\bn
E = \sin{\theta_0} \Pi_0 = \pm \sin^{2}{\theta_0}\sqrt{\frac{1}{|\cos{\theta_0}|}},  \label{clen1}
\en
%The semiclassical spectrum thus obtained is plotted in figure 1. 

The angular momentum  $J_z$ for this class of solution is
$
\frac{J_z}{S}  = \frac{\Pi_0^2}{2}+\cos{\theta_0}-\frac{{R_0}^2}{2}
$
which using the Bohr-Sommerfeld quantization condition eqn.(\ref{quant3}),
\bn
\frac{J_z}{S}= \left(\frac{n}{S}-1\right) -\frac{{R_0}^2}{2}\label{ang_mom_1}
\en
where $R_0^{2}=R_iR_i$. We demand ($J_z-S$) to take integer values. 
This constraints $\frac{R_0^{2}}{2}=\frac{k}{S}$ where $k$ is an integer.
The quantum states corresponding to  the different values of $k$ are
degenerate, consistent with the infinite degeneracy of each Landau level.

For large $n$ limit, $\cos\theta \sim\frac{1}{2n/S}$. The energy of the particle 
in this limit is:
\bn
E= \pm \sqrt{2n/S} \label{largen_sc}
\en

The energy spectrum for different values of $S$ both for 
small values of $n$ and in the asymptotic limit are plotted and discussed in section \ref{results}.

\subsubsection{ \bf{$\vec{n} \perp \mathbf{\Pi}$}}
\label{nperp}
For this ansatz, the components of $\mathbf{\Pi}$ are related to the 
$azimuthal$ angle $\phi$ in the $x-y$ plane as:
\bn
\Pi_x = -\Pi \sin{\phi};\;\;\;
\Pi_y = \Pi \cos{\phi}
\en
so that the energy of the particle, given as :
\bn
E=n_x\Pi_y+n_y\Pi_x =0 \label{clen2}
\en
However, with this choice of co-ordinate axis, 
the variation of the $polar$ angle 
$\theta$ crosses singular points at poles. We choose a 
suitable co-ordinate frame to avoid the singular points.
In the new frame the equations of motion are:
\bea
&&\dot{\Pi}_x = - \cos({\theta});\;\;\; \dot{\Pi}_y =  \sin({\theta}) \sin({\phi})\nonumber \\
&&\dot{\theta} = \Pi_x; \;\;\; \dot{\phi} = \Pi_y - \Pi_x \frac{\cos({\theta})}{\sin({\theta})}\sin{(\phi)} \label{eomscmag2}
\eea
In the $z-x$ plane ($\theta=\pi/2$) the above equations simplifies to:
\bea
\dot{\Pi} = \sin({\phi}); \;\;\; \dot{\phi} = \Pi \label{eomscmag3}
\eea
where $\Pi=\Pi_y$. Combining the above equations we get,
\bn
 \ddot{\phi} + \sin{(\phi)} =0 \label{eqn4phi}
\en
which maps to the pendulum equation. $\phi$ maps to the 
angular co-ordinate or the amplitude, $\Pi$ to the momentum and 
$\cos{\phi}$ to the potential energy of the pendulum. The angular momentum 
\bea
\frac{J_z}{S} &=& \cos{\phi} + \frac{\Pi^{2}}{2} 
\eea
 maps to the total energy $\cos{\phi} + \frac{\dot{\phi}^{2}}{2}$ 
of the pendulum which is a conserved quantity.  
The general solution to eqn.(\ref{eqn4phi}) is Legendre's elliptic function 
of the first kind. $\Pi$ can also be subsequently computed from the values of $\phi$ after
subsequent substitution in eqn (\ref{eomscmag3}). 

For $J_z < S$ the particle will execute $libration$. 
This involves a back-and-forth motion in real space and the 
phase space motion is contractable to a point in topological 
sense. The geometrical phase for this motion is zero. 
For $J_z > S$ the particle will undergo complete $rotations$ in the $z-x$ plane. 
The geometric phase picked up during one complete $rotation$ is $2\pi$.
Since, $\Pi_x=0$, the Bohr-Sommerfeld quantization condition for $J_z>S$ is:  
\begin{eqnarray}
\Omega &=& 2\pi \frac{m}{S}~\Rightarrow S=m 
\end{eqnarray}
where $m$ is an integer.

The zero energy modes for the regime 
$J_z<S$ occurs both for the integer and 
half integer $S$. However as the $J_z$ value increases and $J_z>S$, 
the above Bohr-Sommerfeld quantization condition dictates that the zero modes, 
will appear only for the integer spins.

To summarize, the semiclassical analysis predicts that there will be
a zero energy state for every value of $J_z$ for integer $S$ but for
half odd-integer $S$, they will be zero energy states only for $J_z<S$.

\subsection{Quantum spectrum}
\label{QMag}
In this section, we study the quantum mechanical spectrum of 
the system in a constant external magnetic field $B$. 
In the dimensionless units defined earlier, the hamiltonian 
for the system is:
\bn
\mathcal{H}=\frac{1}{S} \mathbf{S\cdot\Pi}  \label{h3}
\en
with $\left[\Pi_i,\Pi_j\right]=\epsilon_{ij}$.
We define the operators $a$ and $a^{\dagger}$ as,
\bea
a&=&\frac{1}{\sqrt 2}\left(\Pi_x-i\Pi_y\right)\label{adef}\\
a^\dagger &=&\frac{1}{\sqrt 2}\left(\Pi_x+i\Pi_y\right)\label{aadag}
\eea  
they satisfy the commutation relation $[a,a^{\dagger}]=1$.
With the standard notation, $S^{\pm}=S_x \pm i S_y $, the
hamiltonian is: 
\bn
\mathcal{H}= \frac{1}{S\sqrt{2}}\left(S^{-}a^{+}+S^{+}a\right) \label{hqgen}
\en

For $S=1/2$ our model maps on to the strong coupling limit 
of Jaynes-Cummings model\cite{ref11}. For general $S$, 
the model maps to the strong coupling limit of the Tavis-Cummings 
model \cite{ref18, ref19}.   

The Tavis-Cummings model is exactly solved using the Bethe ansatz. However, we 
can deduce several qualitative features of the spectrum from the
symmetry properties alone. The hamiltonian is rotationally invariant. The 
generator of rotations is,
\bn
J_z=a^\dagger a+S_z \label{j3def}
\en
$J_z$ commutes with the hamiltonian and hence in the basis of the eigenstates
of $J_z$, the hamiltonian will be block diagonal. The eigenstates of $J_z$ can
be constructed as follows. Let
\bea
a^\dagger a\vert n,m\rangle&=&\vert n,m\rangle~~~~,n=0,1,\dots\\
S_z\vert n,m\rangle&=&m\vert n,m\rangle~~,m=-S,\dots,S
\eea
we then have,
\bn
J_z\vert n-k,-S+k\rangle=j\vert n-k,-S+k\rangle~~~,j\equiv n-S\label{jzes}
\en
For $-S\le j\le S$, there are $n+1~(=J+S+1)$ values of $k$ and thus the 
corresponding blocks of the hamiltonian will be $J+S+1$ dimensional.
For $J\ge S$ the dimension of the blocks is $2S+1$.

For all values of $S$, there is an operator $\Sigma\equiv e^{i\pi S_z}$
which anti-commutes with the hamiltonian. This implies that the eigenvalues
of $\mathcal{H}$ have to come in pairs, namely if $\epsilon$ is an eigenvalue
then $-\epsilon$ is also an eigenvalue. $\Sigma$ commutes 
with $J_z$ and hence the eigenvalues of each block will have to come in 
pairs. Thus every block with odd dimension has to have at least one 
zero eigenvalue.

The lowest value of $j$ is $j=-S$. There is only one state in this block,
$\vert 0,-S\rangle$. It is easy to verify that this is always an zero energy
eigenstate of $\mathcal{H}$. Thus the blocks $j=-S+2n$ will all have atleast
one zero eigenvalue for $2n<S$. For $j\ge S$, the dimension of the blocks, as
mentioned earlier, is $2S+1$. Thus for integer $S$, there is atleast one
zero eigenvalue in each such block. For half-odd integer $S$, this is not
necessary.

The hamiltonian can also be diagonalised analytically in the $j>>S$ limit.
In this limit, the hamiltonian reduces to
\bn
H=\frac{\sqrt{2n}}{S} S^y\left(1+o(1/n)\right)
\en
where $n=J+S$. The energy eigenvalues of the $2S+1$ branches, in this limit 
are therefore,
\bn
E=\sqrt{2n}\frac{m}{S},~~m=-S,\dots,S \label{largen_q}
\en

\begin{figure*}[htp]

\subfloat{\label{compareplot1}\includegraphics[width=.50\textwidth]{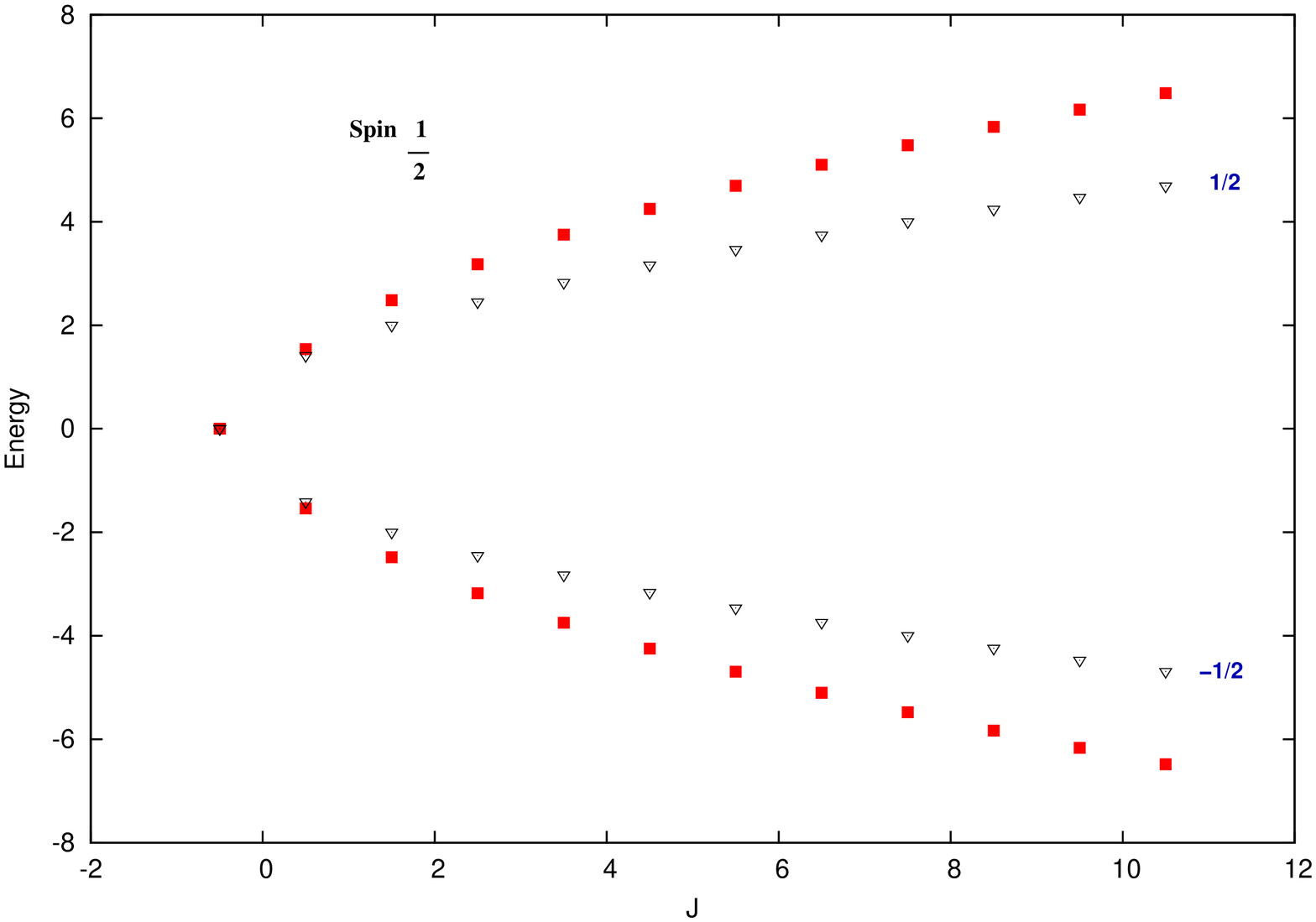}}
\subfloat{\label{compareplot2}\includegraphics[width=.50\textwidth]{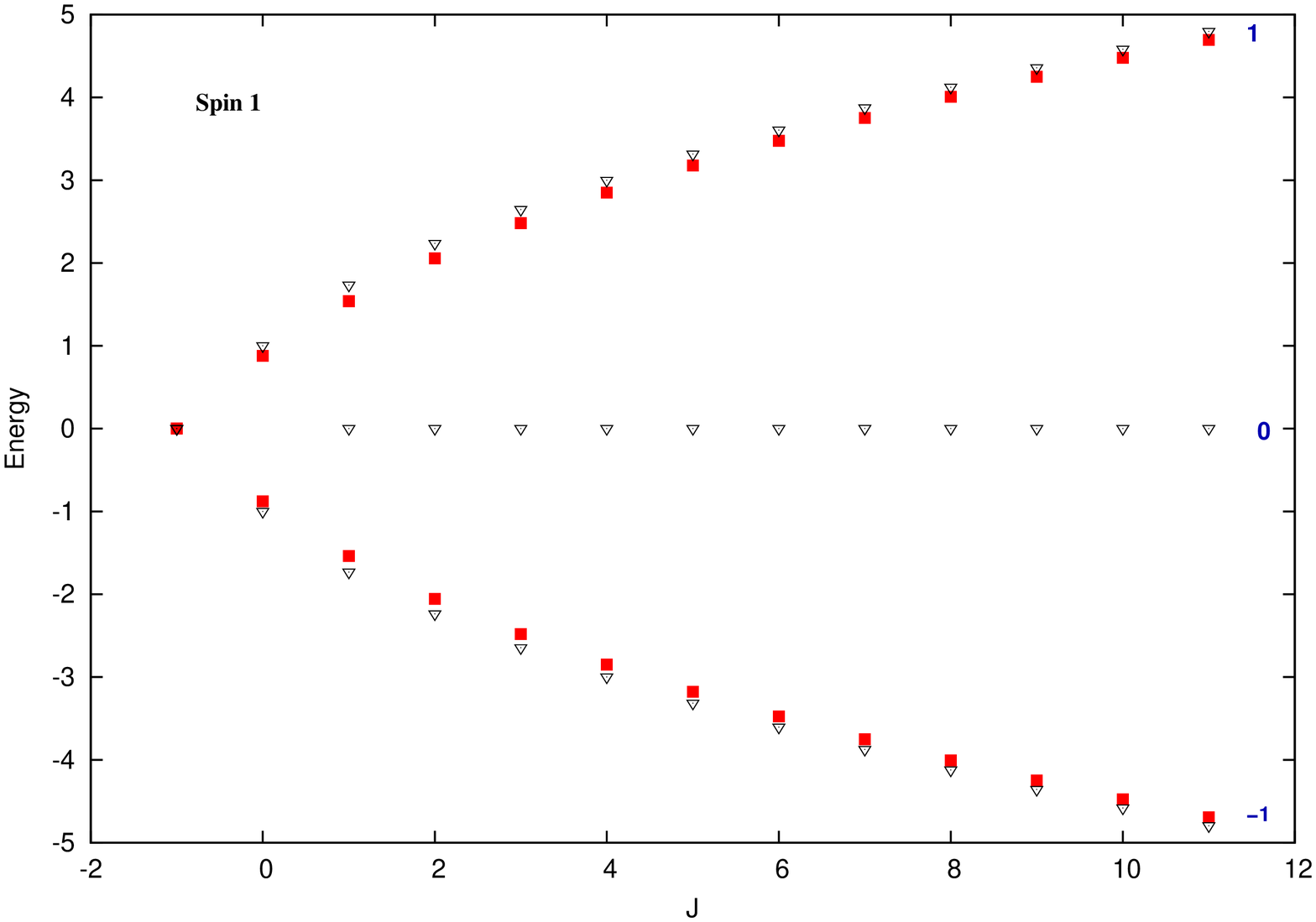}}\\
\subfloat{\label{compareplot4}\includegraphics[width=.50\textwidth]{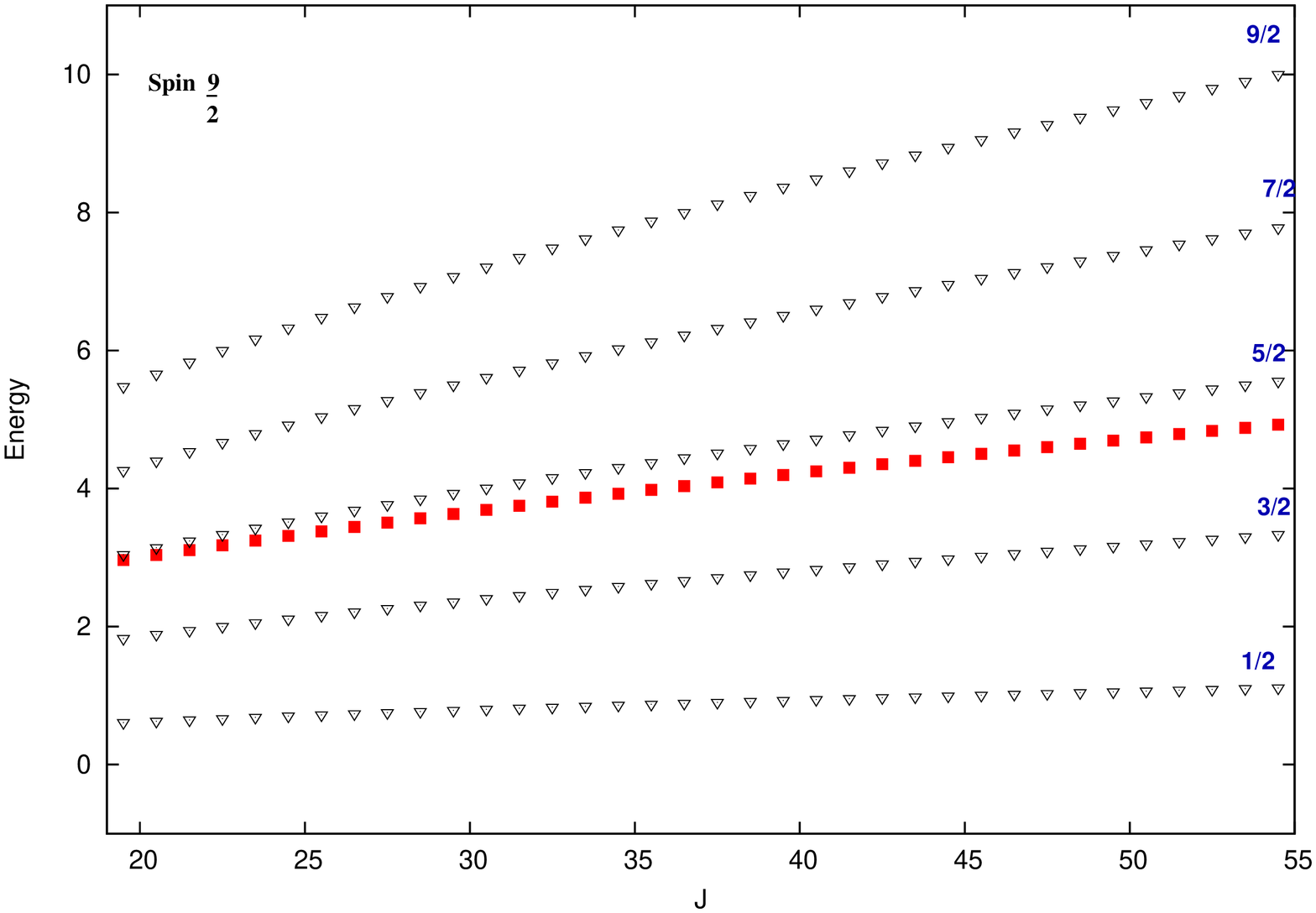}}
\subfloat{\label{compareplot5}\includegraphics[width=.50\textwidth]{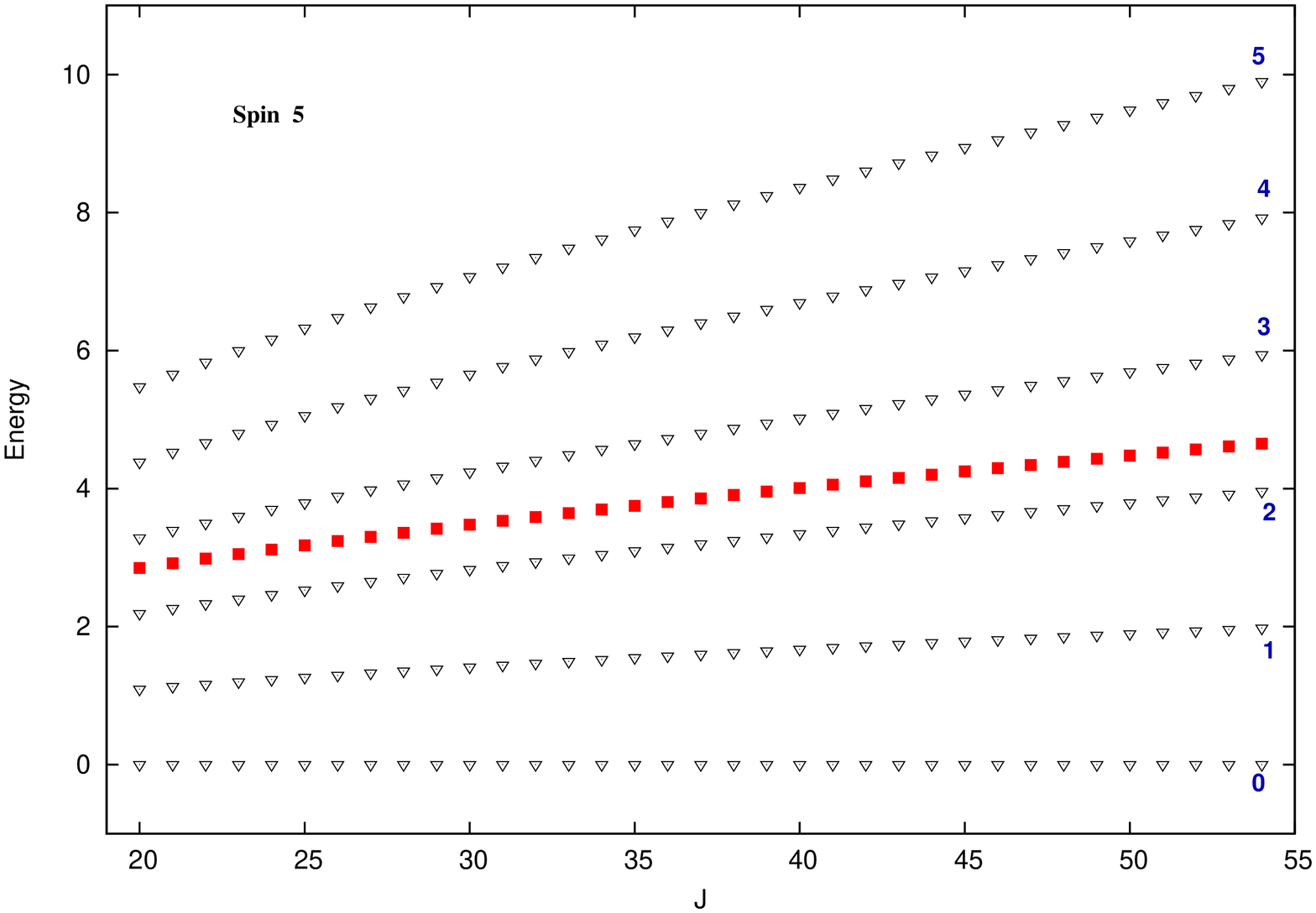}}

%\captionsetup{singlelinecheck=off}

\caption{ (Color online) Plots showing the energy spectrum obtained from semi-classical 
analysis and the exact quantum spectrum. The curves with red square points are obtained 
from semi-classical analysis. The quantum spectrum is denoted by black triangles. 
Figs. (\ref{compareplot1}),(\ref{compareplot2}), (\ref{compareplot4}),(\ref{compareplot5}) 
are the plots for $S= $1/2, 1, 9/2 and 5 respectively. For the quantum spectrum, the energy branches 
corresponding to the different \textit{$S_z$} values for a given \textit{S} are marked. To 
avoid clutter only the positive energy branches of the spectrum for high $S$ values are plotted. 
In the large \textit{n} limit, the semi-classical spectrum gives the $\sqrt{S} -th$ branch of the quantum spectrum.
} 
\label{compareplots}
\end{figure*}

The energy eigenvalues can be numerically computed for low values 
of $n$. We have shown the spectrum for $S=\frac{1}{2},1$
and $S=\frac{9}{2},5$ in figure [\ref{compareplots}].

\section{Discussion}

\label{discussion}
We now compare the results of the semiclassical quantization and the 
exact spectrum. In the free particle case the two spectra agree exactly.
There is no exact match of the spectrum in a constant magnetic field and
comparison is more interesting.

We first consider the zero-energy states. The semi-classical analysis
predicts that zero energy modes will occur for $j<S$. This is consistent
with the results obtained from the exact analysis. The quantum
spectrum has zero modes only the odd dimensional subspaces. This detail
is not reproduced by the semiclassical analysis. However, we note that
that the semiclassical regime in our formalism is the opposite limit,
$j>S$. In this regime the semiclassical analysis predicts zero modes
only for integer $S$ and that is indeed bourne out by the exact spectrum.

The exact spectrum has $2S+1$ branches for every $j>S$. However, we have only
two branches of classical solutions. So which branch does it correspond to ?
A comparison of the asymptotic results for the exact and the semiclassical
asymptotic analysis, equations (\ref{largen_q}) and (\ref{largen_sc}) shows that the 
classical solutions correspond to $m=\pm\sqrt{S}$. Namely, should lie between $\pm m$ and 
$\pm(m+1)$ where $m<\sqrt{S}<m+1$. Our numerical computations plotted in figure
\ref{compareplots} shows that this is true even for small values of $S$. 
For $S=1/2$, while the $\sqrt{2n}$ dependence of the spectrum is reproduced,
the semiclassical spectrum overestimates the energies by a factor 
$\approx\sqrt{2}$.

\section{Results and conclusions}
\label{results}
To summarize our results, the semiclassical spectrum obtained by 
Bohr-Sommerfeld quantization reproduces the exact quantum spectrum 
for the case of a free particle on a cylinder.

For the particle in a constant magnetic field, the semiclassical
spectrum correctly predicts the occurrence of zero energy branches
for integer $S$ and their absence for half-odd integer $S$ for $j>S$.
The exact solutions have $2S+1$ branches in this regime. The classical
solutions we have found yields a average of these branches. We note
that the level spacing between these branches is $\sim S^0$. Hence 
more detailed quantitative agreement may be obtained by doing the gaussian
fluctuations about the classical solutions instead of Bohr-Sommerfeld
quantization.

%\section{Acknowledgements}

We thank M.V.N. Murthy, Matthias Brack, Diptiman Sen and Bindu Bambah for 
useful discussions. M.M thanks M. Deka for discussions related to the numerical 
computation.


\section*{References}
\begin{thebibliography}{99}

\bibitem{ref1}  Novoselov K S, Geim A K ,  Morozov S V, Jiang D , Zhang Y , 
 Dubonos S V,  Grigorieva I V, and  Firsov A A 2004 \textit{Science}, {\bf 306}, 666; 
 Novoselov K S,  Geim A K,  Morozov S V,
 Jiang D,  Katsnelson M I,  Grigorieva I V,  Dubonos S V,
and  Firsov A A 2005 , \textit{Nature}, {\bf 438}, 197; 
 Zhang Y, Tan Y W,  Stormer H L and  Kim P 2005 \textit{Nature}, {\bf 438}, 201.

 \bibitem{ref2}  Bernevig B A and  Zhang S C 2006 \textit{Phys. Rev. Lett.} 
 \textbf{96}, 106802; Bernevig B A, Hughes T L and  Zhang S C 2006 \textit{Science} 
 \textbf{314}, 1757;  K$\rm{\ddot{o}}$nig M,  Wiedmann S,  Br\"{u}ne C,  Roth A, 
 Buhmann H,  Molenkamp L W,  Qi X L and  Zhang S C 2007 \textit{Science} 
 \textbf{318}, 766;  Hassan M Z and  Kane C L 2010 \textit{Rev. Mod. Phys.}
\textbf{82}, 3045 and the references therein.

\bibitem{ref3}  Fu L and  Kane C L 2008 \textit{Phys.Rev. Lett.},\textbf{100}, 096407;
  Akhmerov A R,  Nilsson J and  Beenakker C W J 2009 \textit{Phys.Rev. Lett.},\textbf{102}, 216404;
 Tanaka Y,  Yokoyama T and Nagaosa N 2009 \textit{Phys. Rev. Lett.}, \textbf{103}, 107002.

\bibitem{ref4}  Mondal S,  Sen D,  Sengupta K and  Shankar R 2010 \textit{Phys. Rev. Lett.}, \textbf{104}, 046403,
 Mondal S,  Sen D,  Sengupta K and  Shankar R 2010 \textit{Phys. Rev. B}, \textbf{82}, 045120.

\bibitem{ref5}  Z\"{u}licke U,  Bolte J and  Winkler R 2007 \textit{New Journal of Physics},  \textbf{9} 355.

 \bibitem{ref6}  Rusin T M and  Zawadzki W 2008 \textit{ Phys. Rev. B}, \textbf{78}, 125419.

 \bibitem{ref7}  Zawadzki W 2011 \textit{J. Phys.: Condens. Matter}, \textbf{23}, 143201.

\bibitem{yip}  Yip P 1983 \textit{ J. Math. Phys.}, \textbf{24}, 1206.

\bibitem{ref8} Chen Y L, Analytis J G, Chu J H, Liu Z K, Mo S K, Qi X L, Zhang H J,
Lu D H, Dai X, Fang Z, Zhang S C, Fisher I R, Hussain Z, Shen Z X and Shen Z X 2009
\textit{Science} \textbf{325}, 178; Xia Y, Qian D, Hsieh D, Wray L, Pal A, Lin H,
Bansil A, Grauer D, Hor Y S, Cava R J and Hasan M Z 2009 \textit{Nature Phys.}
\textbf{5}, 398.
  

 \bibitem{ref9}  Pauli W 1932 \textit{Helv. Phys. Acta} \textbf{5}, 179.

 \bibitem{ref10}  Broglie L de, \textit{La Th\'{e}orie des Particules de Spin 1/2} 
 (Gauthier-Villars, Paris, 1952).

 \bibitem{ref11}  Rubinow S I and  Keller J B 1963 \textit{Phys. Rev.} \textbf{1}, 2789.

 \bibitem{ref12}  Yabana K and  Horuichi H 1987 \textit{Prog. Theor. Phys.} \textbf {77}, 517.

 \bibitem{ref13}  Kuratsuji H and  Iida S 1985 \textit{Prog. Theor. Phys.} \textbf {74}, 439.

 \bibitem{ref14}  Kuratsuji H and  Iida S 1988 \textit{Phys. Rev. D} \textbf {37}, 441.

 \bibitem{ref15}  Bolte J and  Keppeler S 1998 \textit{Phys. Rev. Lett.} \textbf{81}, 1987 ; 
                Bolte J and  Keppeler S  1999 \textit{Ann. Phys., NY}  \textbf{274}, 125.

 \bibitem{ref16}  Pletyukhov M and  Zaitsev O 2003 \textit{J. Phys. A: Math. Gen.} \textbf{36}, 5181.

 \bibitem{ref17}  Pletyukhov M,  Amann C,  Mehta M and  Brack M 2002 \textit{Phys. Rev. Lett.} \textbf{89}, 116601.

 \bibitem{ref18}  Bogoliubov N M,  Bullough R K and  Timonen J 1996 \textit{J. Phys. A: Math. Gen.}, \textbf{29} 6305 .

 \bibitem{ref19}  Lee Y H,  Links J and  Zhang Y Z 2011 \textit {Nonlinearity}, \textbf{24}, 1975 .

\end{thebibliography}
\end{document}